\documentclass[prl,aps,preprint,showpacs]{revtex4-1}
\usepackage{graphicx}

\begin{document}
\title{Comment on ``Tip induced unconventional superconductivity on Weyl semimetal TaAs" [arXiv:1607.00513]}
\author{Sirshendu Gayen}
\author{Leena Aggarwal}
\author{Goutam Sheet}
\email {goutam@iisermohali.ac.in}
\affiliation{Department of Physical Sciences,
Indian Institute of Science Education and Research Mohali,
Sector 81, S. A. S. Nagar, Manauli, PO: 140306, India}
\date{\today}

\begin{abstract}
Recently, Wang $et$ $al.$ have reported the observation of ``unconventional superconductivity" in the Weyl semimetal TaAs [arXiv:1607.00513]. The authors have written ``\textit{A  conductance  plateau  and  sharp double dips  are  observed  in  the  point  contact spectra, indicating p-wave   like   unconventional   superconductivity. Furthermore, the   zero   bias conductance peak in low temperature regime is detected, suggesting potentially the existence of Majorana zero modes. The experimentally observed tunneling spectra can be interpreted with a novel mirror-symmetry protected topological superconductor induced in TaAs, which can exhibit zero bias and double finite bias peaks, and double conductance dips in the measurements}." It is known that for a superconducting point contact, the features like a zero-bias conductance peak, a plateau and single or multiple conductance dips might arise due to simple contact-heating related effects. Such features are routinely observed in point contacts involving a wide variety of superconductors including simple conventional superconductors like Nb, Pb and Ta, when the experiments are not performed in the right regime of mesoscopic transport and such spectra do not provide any energy resolved spectroscopic information. Here we show that the data presented by Wang $et$ $al.$ do not confirm the existence of ``superconductivity" and discuss how a tip-induced superconducting (TISC) phase can be confirmed by performing measurements in different regimes of transport. Even if it is assumed that Wang $et$ $al.$ achieved a TISC phase on TaAs, all the spectra that they have reported show striking similarities with the type of spectra expected in thermal regime of transport involving superconducting point contacts. Such data cannot be used for extracting any spectroscopic information and based on such data any discussion on ``p-wave" superconductivity or the emergence of Majorana modes should be considered invalid. This version (v2) also includes a brief discussion on the response of Wang $et$  $al.$ [arXiv:1607.02886 (2016)] to the first version (v1) of this comment. Correct ballistic regime data on TaAs point contacts can be found in arXiv:1607.05131 (2016).

\end{abstract}

\maketitle

Point contact spectroscopy is a powerful technique which has helped the community understand the Fermi surface properties of a large number of materials with great success for several decades.\cite{naidyuk} At the same time, it is also known that like all other experimental techniques, point contact spectroscopy has its own limitations.\cite{prb04} When the limitations of the technique are not clearly understood there is a risk of generating misleading results -- such results, sometimes published in well reputed journals, confuse the physics community in general.\cite{Comment_Laura, prl06} 

In this comment we focus on a recent paper by Wang $et$ $al.$ on TaAs.\cite{Wang_TaAs} It is well known that it is possible to obtain a tip-induced superconducting (we named it ``TISC") phase in topologically non-trivial materials under mesoscopic point contacts. This was first shown by Aggarwal $et$ $al.$ in arXiv:1410.2072 where TISC was shown for the first time on polycrystalline samples of Cd$_{3}$As$_{2}$ where the grain size was shown to be large enough so that the point contacts could be made majority of the times on individual single crystalline grains. Subsequently, after 3 months, some of the authors of the paper being commented on here reproduced the TISC phase on single crystalline Cd$_{3}$As$_{2}$ (arXiv:1501.00418). Eventually, both the papers simultaneously appeared in Nature Materials, but the precedence of the paper by Aggarwal $et$ $al.$ established in arXiv.org is also reflected in the receipt dates in the journal. Unfortunately, this simultaneity of publication has been emphasized in the reply\cite{Wang} to our comment in v1.\cite{Goutam} The remark on sample quality in the reply of Wang $et$ $al.$ is to be understood as essential to a follow-up experiment. 

Despite several theoretical attempts, the origin of such a TISC phase is still not understood. The TISC phase remains elusive mainly because the phase appears only under point contacts where bulk characterization techniques fail. Andreev reflection spectroscopy of such TISC comes as a rescue. However, it must be noted that Andreev reflection at a point contact can be used as a spectroscopic probe only when the point contacts are made in the ballistic or in the diffusive regimes of transport. When the point contact experiments are performed away from the ballistic regime, depending on the geometry of the point contacts, the spectrum might have multiple sharp features (artefacts), often symmetric about $V = 0$. 

The point contact spectra presented by Wang $et$ $al.$ in Figure 1 of Ref. \cite{Wang_TaAs} show two features based on which the authors have inferred the ``unconventional" nature of ``superconductivity" -- (a) sharp conductance dips at high bias and (b) a conductance peak/plateau at zero bias. Such conductance dips are ubiquitously observed for point contacts between a wide variety of superconductors and normal metals. A number of such spectra involving conventional superconductors like Nb and Ta were shown by Sheet $et$ $al.$ where such conductance dips were shown to originate from the critical current of the superconducting point contacts when the point contacts were not in the ballistic regime of transport\cite{prb04}. For such point contacts in the thermal regime, multiple conductance dips might also arise depending on number of distinct electrical contacts established. Here we show some representative spectra where features similar to that obtained on TaAs by Wang $et$ $al.$ can be seen in point contacts with elemental superconductors like Nb and Pb. It should be noted that the spectra obtained in the thermal limit of transport may show arbitrary shapes depending on the contact geometry – it can be ``U"-shaped, ``V"-shaped, ``plateau"-like etc.

\begin{figure}[h!]
	\centering
		\includegraphics[width=\textwidth]{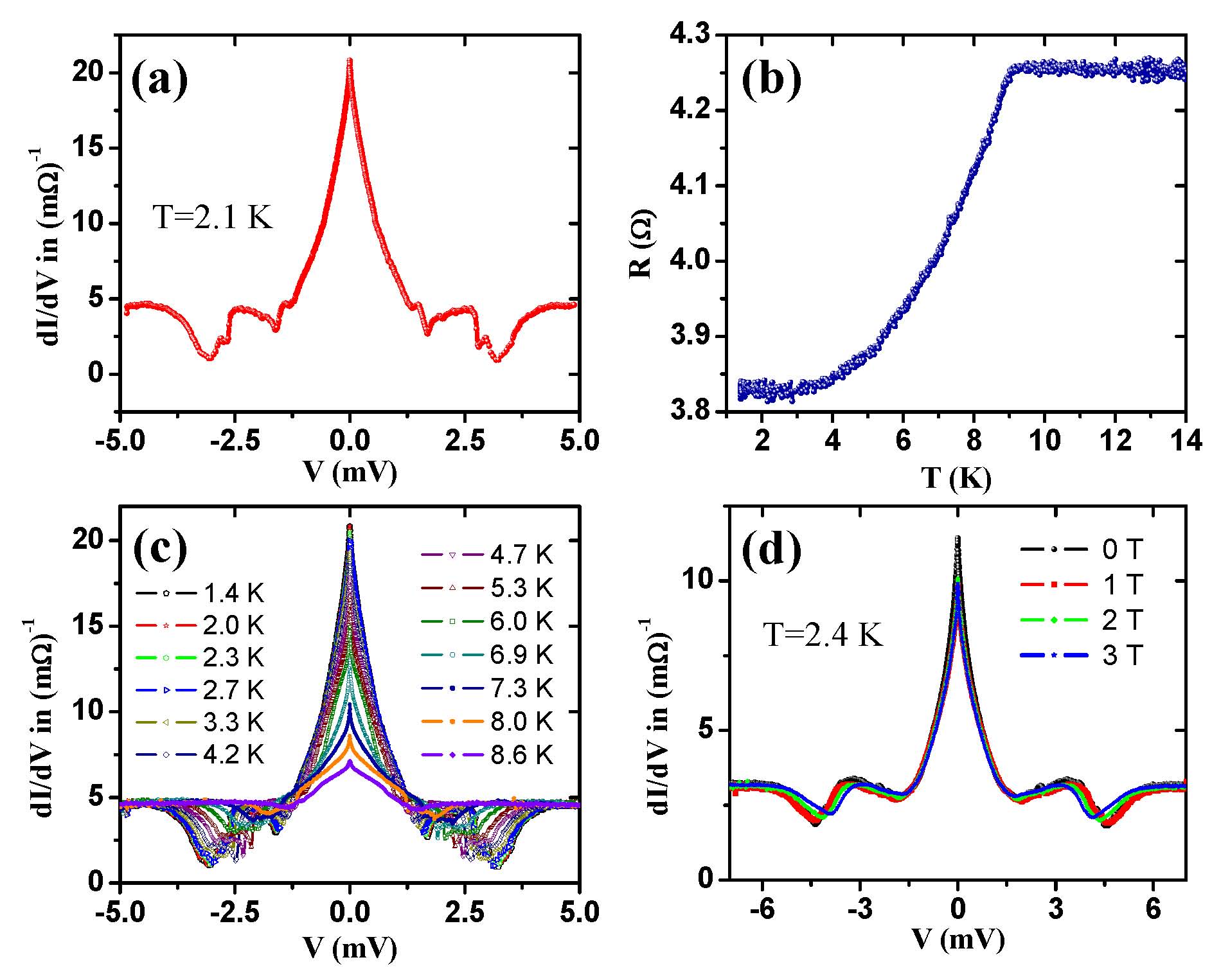}
	\caption{(a) Point contact spectrum obtained on superconducting Nb with a Au tip showing multiple conductance dips. (b) Temperature dependence of resistivity of Nb/Au point contact at zero magnetic field. (c) Systematic temperature Variation of point contact spectra. (d) Point contact spectra under magnetic field applied. Similarities of the data obtained on the conventional superconductor Nb with those obtained on TaAs as reported by Wang $et$ $al.$ should be noted. }
	\label{f1}
\end{figure}

In Fig. 1(a) we show a point contact spectrum obtained on superconducting Nb with a Au tip. Multiple conductance dips are clearly observed and are indicated by arrows in the figure. In Fig. 1(b) we show the resistive transition observed in the temperature dependence of the point contact resistance confirming that the point contacts are in the thermal regime of transport. As expected for superconducting point contacts, all the high-bias conductance dips and the zero-bias conductance peak evolve systematically with increasing temperature (Fig. 1(c)) and magnetic field (Fig. 1(d)) respectively. In Fig. 2 we show four representative spectra between the conventional superconductor Pb and Ag where multiple critical current dominated conductance dips followed by a zero-bias conductance peak/plateau are clearly seen. This set of data shows when the point contacts are not in the ballistic regime, a large number of spectral shapes are expected, even on simple well known conventional superconductors, primarily due to contact-heating dominated artefacts. Striking similarities of some of these artefact-dominated spectra with the spectra presented by Wang $et$ $al.$ can be seen. However, such data do not give any information about conventionality or unconventionality of a superconductor.

\begin{figure}[h]
	\centering
		\includegraphics[width=0.8\textwidth]{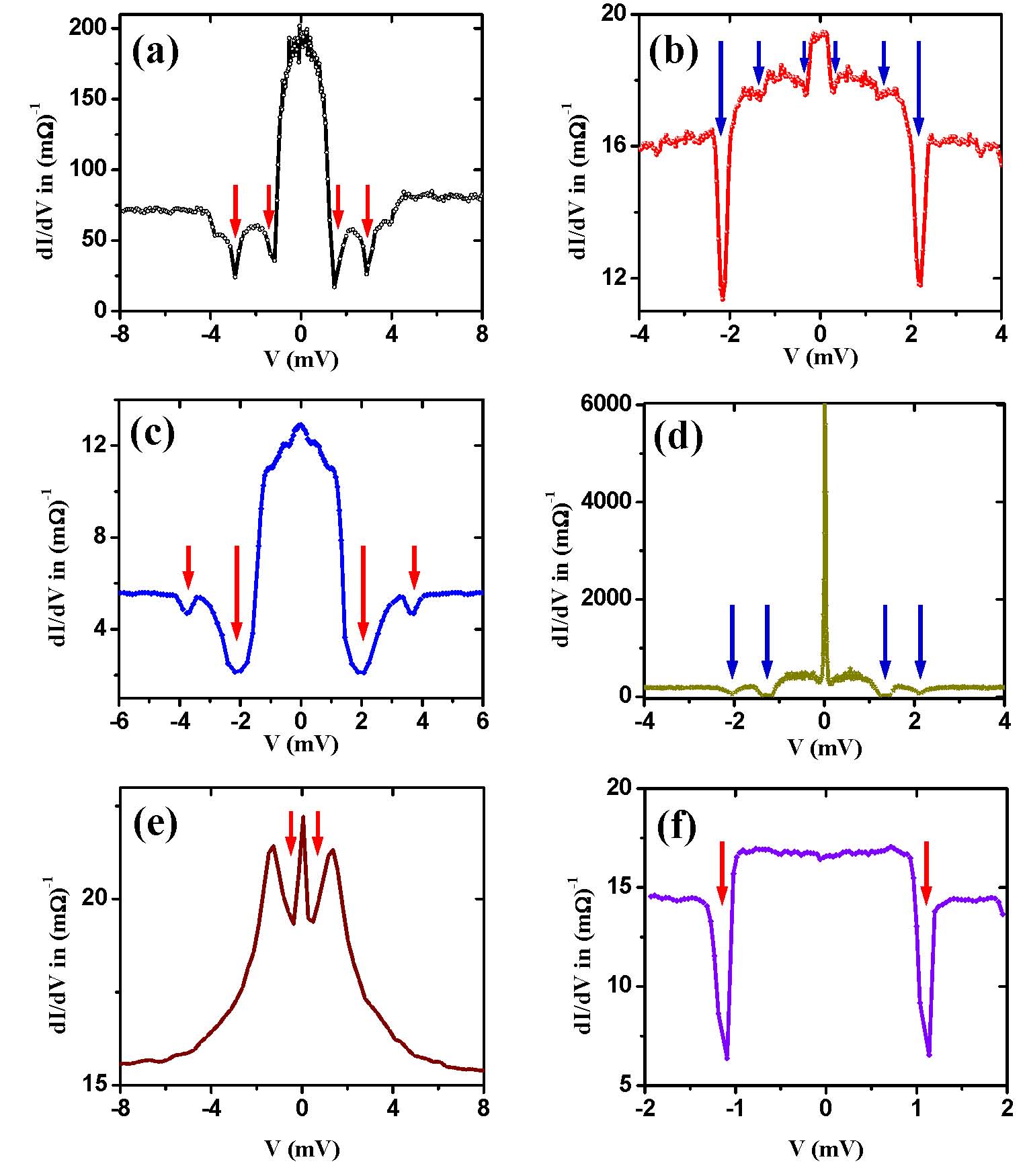}
	\caption{(a)-(f) Point contact spectra with distinct (multiple) conductance dips obtained on superconducting Pb with a Ag tip, when the point contacts are away from the ballistic regime. Most of the spectra with the conventional superconductor Pb show striking similarities with those obtained on TaAs as reported by Wang $et$ $al.$ In (f) a clear conductance plateau followed by conductance dips is also observed.}
	\label{f2}
\end{figure}

In fact, Wang $et$ $al.$ has provided a discussion on the possibility of critical current related effects in the supplemental material. From the analysis of the point contact resistance in presence (and absence) of magnetic fields they have attempted to prove that the role of critical current is zero in their spectra because their estimate of the Maxwell's resistance is negligible. Here we show that their proposed analysis for calculating the Maxwell's resistance is erroneous. First, they have not explained why they believe that the assumption (in their supplementary materials)  $R_{PC} = R_{Sh} = 18.8 \Omega$ is valid. It is logically inconsistent to assume the point contact to be ballistic in order to prove it to be ballistic. Second, when a topologically non-trivial system is involved in the point contact, the authors have not discussed why they believed that the magnetic field dependence would originate only from the sample. Rationally, the point contact itself should have large magnetoresistance, particularly because the point contact is a completely different phase (even possibly superconducting). Third, the authors started their discussion saying that the point contact is ballistic only when the contact size is less than the normal state mean free path of the sample. However, they have not provided such a comparison. Even for point contacts with finite Z, the point contact diameter can be calculated using Wexler’s formula and then the same can be compared with the measured mean free path. In short, the analysis presented by Wang $et$ $al.$ to find out the Maxwell's contribution is non-trivial, erroneous and to the knowledge of the authors of this comment, was never used for testing the ballisticity of point contacts. Furthermore, the thermal limit data shown in Fig. 1 and Fig. 2 of this comment have been obtained using pure metallic tips and foils. If the scheme of Wang $et$ $al.$ is followed for such point contacts between two pure elements, a very small value of Maxwell's resistance is expected based on which the non-ballistic nature of these point contacts cannot be ruled out.

Such a non-trivial analysis, however, is not required if the data presented by Wang $et$ $al.$ are considered carefully. Wang $et$ $al.$ have shown a transition in the point contact resistance (Figure 2(a) and inset of Figure 3(b)) which they have claimed to be a signature of superconducting transition. If the point contact is in the ballistic regime, the bulk resistivity does not contribute to the point contact resistance. In that case, it is not clear how can one expect to observe the resistive transition in the ballistic point contact resistance! 
Wang $et$ $al.$ has shown, in their reply to the previous version of our comment, how BTK theory predicts the temperature dependence of the zero-bias resistance of ballistic point contacts due to the suppression of Andreev reflection with increasing temperature. It should be noted that before claiming the observation of Andreev reflection, the existence of superconductivity must be proved beyond doubt. In their paper Wang $et$ $al.$ used the temperature dependence of point contact resistance as a proof of superconductivity. We wonder how can the same data be used for proving “Andreev reflection”! We did not find a single data in their paper that even remotely matches with Andreev reflection dominated spectra as predicted by BTK theory. Therefore, the temperature dependence presented by Wang $et$ $al.$ may emerge from a number of other phenomena, than superconductivity, where the zero-bias density of states decreases with increasing temperature. In their Cd$_{3}$As$_{2}$ paper Aggarwal $et$ $al.$ had clearly demonstrated how a TISC phase can be confirmed beyond doubt – i.e., by exploring different regimes of mesoscopic transport and demonstrating the hallmark signatures of superconductivity in those regimes. This scheme has been discussed in detail by Das $et$ $al.$ in other papers like in arXiv:1607.01609 (2016)\cite{Shekhar} and arXiv:1607.05131 (2016)\cite{taas}.  In these papers the authors have provided a checklist for TISC without which the phase cannot be confirmed beyond ambiguity. Without such rigorous experimental evidence every claim of TISC must be questioned. 
Within the formalism of BTK theory a conductance plateau is observed only at zero temperature and for Z = 0. Both of these are ideal conditions that cannot be achieved in real experiments. Z = 0 is usually not achieved in reality because there is always some amount of Fermi velocity mismatch between two different materials forming a point contact. However, as we have shown in our comment, when the superconducting point contacts are made in the thermal regime of transport, all kinds of spectral features, ``V"-shaped, ``U"-shaped, ``plateau"-shaped can be obtained. A long list of such spectra are provided in several published literature. Such spectral features may or may not be accompanied by single or multiple dips. Such data give no information about the nature of superconductivity – conventional/unconventional. Without taking these effects into consideration, we believe, any claim of TISC, conventional or not, should be considered invalid. 
Furthermore, in the title of their paper they claim observation of ``unconventional superconductivity" in TaAs point contacts and in their reply they wrote ``our results can be reasonably interpreted by the BTK model with a finite barrier".  These two claims are clearly contradictory as BTK theory is valid only for conventional superconductors. Furthermore, BTK theory never predicts conductance dips, single or multiple and hence it is not clear how their results can be “reasonably interpreted” by BTK theory.

Therefore, as per the discussion presented above, the claim of TISC in TaAs by Wang $et$ $al.$ is not beyond doubt. If Wang $et$ $al.$ strongly believe that they have probed superconducting point contacts on TaAs in the ballistic regime, we wonder, had they repeated their experiments on TaAs in the thermal or intermediate regime of transport, what kind of spectra would they expect? 

\textbf{The correct ballistic limit data with direct proof of superconductivity in TaAs point contacts and detailed analysis based on well understood theoretical concepts can be found in arXiv:1607.05131 (2016).\cite{taas}}

We thank Jithin Bhagwathi and Preetha Saha for their help during some of the point contact experiments presented in this comment. We also thank Professor Praveen Chaddah for reviewing the text of our revised comment and for his extremely useful suggestions.


\begin{thebibliography}{100}  

\bibitem{naidyuk} Y. G. Naidyuk, I. K.Yanson, \textit{Point-contact Spectroscopy}, Springer (2005).

\bibitem{prb04} G. Sheet, S. Mukhopadhyay, P. Raychaudhuri, \textit{Role of critical current on the point-contact Andreev reflection spectra
between a normal metal and a superconductor} Phys. Rev. B \textbf{69}, 134507 (2004).

\bibitem{prl06} G. Sheet, P. Raychaudhuri, \textit{Comment on ``Spectroscopic Evidence for Multiple Order Parameters in the Heavy Fermion Superconductor CeCoIn$_5$''} Phys. Rev. Lett. 96, 259701 (2006). 

\bibitem{Comment_Laura} W. K. Park and L. H. Greene, \textit{Comment on ``Spectroscopic Evidence for Multiple Order Parameter Components in the Heavy Fermion Superconductor CeCoIn$_5$''} Phys. Rev. Lett. 96, 259702 (2006).

\bibitem{Wang_TaAs} H. Wang, H. Wang, Y. Chen, J. Luo, Z. Yuan, J. Liu, Y. Wang, S. Jia, X.-J. Liu, J. Wei, J. Wang  \textit{Tip induced unconventional superconductivity on Weyl semimetal TaAs} arXiv:1607.00513 (2016).

\bibitem {Wang} H. Wang, H. Wang, Y. Chen, J. Luo, Z. Yuan, J. Liu, Y. Wang, S. Jia, X.-J. Liu, J. Wei, J. Wang, \textit{Reply to Comment on ``Tip induced unconventional superconductivity on Weyl semimetal TaAs"}. arXiv:1607.02886 (2016).

\bibitem {Goutam} S. Gayen, L. Aggarwal \& G. Sheet, \textit{Comment on "Tip induced unconventional superconductivity on Weyl semimetal TaAs"}. arXiv:1607.01405 (2016).

\bibitem{leena} L. Aggarwal, A. Gaurav, G. S. Thakur, Z. Haque, A. K. Ganguli, G. Sheet, \textit{Unconventional superconductivity at mesoscopic point contacts on the 3D Dirac semimetal Cd$_3$As$_2$}  Nature Materials \textbf{15}, 32 (2016); arXiv:1410.2072 (2014).

\bibitem{jian}  H. Wang, H. Wang, H. Liu, H. Lu, W. Yang, S. Jia, X.-J. Liu, X. C. Xie, J. Wei, J. Wang,	\textit{Observation of superconductivity induced by a point contact on 3D Dirac semimetal Cd$_3$As$_2$ crystals}
Nature Materials \textbf{15}, 38 (2016); arXiv:1501.00418 (2015).

\bibitem{Shekhar} S. Das, L. Aggarwal, S. Roychowdhury, M. Aslam, S. Gayen, K. Biswas and G. Sheet \textit{Unexpected superconductivity at nanoscale junctions made on the topological crystalline insulator Pb$_{0.6}$Sn$_{0.4}$Te} arXiv:1607.01609 (2016).

\bibitem{taas} Leena Aggarwal, Sirshendu Gayen, Shekhar Das, Ritesh Kumar, Vicky S\"{u}{\ss}, Chandra Shekhar, Claudia Felser, Goutam Sheet \textit{Mesoscopic superconductivity and high spin polarization coexisting at metallic point contacts on the Weyl semimetal TaAs} arXiv:1607.05131 (2016).



\end{thebibliography}
\end{document}